\documentclass[
prl,
nofootinbib,
twocolumn,
superscriptaddress,
showpacs
]{revtex4}

\usepackage{graphicx}% Include figure files
\usepackage{color} % use colored fonts
\usepackage{dcolumn}% Align table columns on decimal point
\usepackage{bm}% bold math
\usepackage{epsfig}
\usepackage{latexsym}
\usepackage{amsmath}
\usepackage{amsfonts}
\usepackage{amsxtra}

\newcommand{\lwrsim}{\raise0.3ex\hbox{$<$\kern-0.75em\raise-1.1ex\hbox{$\sim$}}}
\def\krto{ {\,\,\lower .8ex\hbox {$\longrightarrow \atop k \rightarrow 0$}\,\,}}

\def\bea{\begin{eqnarray} }
\def\beq{\begin{eqnarray} }

\def\eea{\end{eqnarray}}
\def\eeq{\end{eqnarray}}

\newcommand{\ams}{\alpha_{\overline{\rm MS}}}
\newcommand{\Lms}{\Lambda_{\overline{\rm MS}}}
\newcommand{\refeq}[1]{Eq.~\eqref{#1}}

\begin{document} %\date{\today}
%\date{}
%%%%%%%%%%%%%%%%%%%%%%%%%%%%%%%%%%%%%%%%%%%%%%
%% title page 
%%%%%%%%%%%%%%%%%%%%%%%%%%%%%%%%%%%%%%%%%%%%%%

\title{The strong running coupling from the gauge sector of Domain Wall lattice QCD with physical quark masses}

\author{S.~Zafeiropoulos}
\affiliation{Institute for Theoretical Physics, Heidelberg University, Philosophenweg 12, 69120 Heidelberg, Germany}
%----
\author{Ph.~Boucaud} 
\affiliation{ Laboratoire de Physique Th\'eorique (UMR8627), CNRS, Univ. Paris-Sud, Universit\'e Paris-Saclay, 
91405 Orsay, France}
%---
\author{F.~De Soto}
\affiliation{Dpto. Sistemas F\'isicos, Qu\'imicos y Naturales, 
Univ. Pablo de Olavide, 41013 Sevilla, Spain}
\affiliation{CAFPE, Universidad de Granada, E-18071 Granada, Spain}
%--
\author{J.~Rodr\'{\i}guez-Quintero}
\affiliation{Department of Integrated Sciences and Center for Advanced Studies in Physics, Mathematics and Computation; 
University of Huelva, E-21071 Huelva; Spain.}
\affiliation{CAFPE, Universidad de Granada, E-18071 Granada, Spain}
%---
\author{J.~Segovia} 
\affiliation{Dpto. Sistemas F\'isicos, Qu\'imicos y Naturales, 
Univ. Pablo de Olavide, 41013 Sevilla, Spain}
\affiliation{CAFPE, Universidad de Granada, E-18071 Granada, Spain}
%-----

\begin{abstract}

We report on the first computation of the strong running coupling at the physical point (physical pion mass) from the ghost-gluon vertex, computed from lattice simulations with three flavors of Domain Wall fermions. We find $\ams(m_Z^2)=0.1172(11)$, in remarkably good agreement with the world-wide average. Our computational bridge to this value is the Taylor-scheme strong coupling, which has been revealed of great interest by itself because it can be directly related to the quark-gluon interaction kernel in continuum approaches to the QCD bound-state problem.

\end{abstract}


\maketitle

%\begin{flushright}
%%DAMTP-2011-nnn\\
%LPT-Orsay 11-74\\
%UHU-FT/11-29 \\
%%LPSC-11-nnn \\
%IRFU-11-136
%\end{flushright}
%%
%\vspace*{-1cm}
%%
%\begin{figure}[h]
%%  \begin{center}
%    \includegraphics[width=25mm]{figs/ETMC_rund.pdf}
%%  \end{center}
%\end{figure}

%\vfill
%\newpage

%%%%%%%%%%%%%%%%%%%%%%%%%%%%%%%%%%%%%%%%%%%%%%%%%%%%%%%%%
%% end of title page
%%%%%%%%%%%%%%%%%%%%%%%%%%%%%%%%%%%%%%%%%%%%%%%%%%%%%%%%%

%%%%%%%%%%%%%%%%%%%%%%%%%%%%%%%%%%%%%%%%%%%%%%%%%%%%%%%%%
%% body of the paper
%%%%%%%%%%%%%%%%%%%%%%%%%%%%%%%%%%%%%%%%%%%%%%%%%%%%%%%%%

%\section{Introduction}
%\alinea

\noindent\emph{Introduction}.\,---\,
Quantum Chromodynamics (QCD), the non-Abelian gauge quantum field theory describing the strong interaction between 
quarks and gluons, can be compactly expressed in one line with a few inputs; namely, the current quark masses and 
the strong coupling constant, $\alpha_s$~\cite{Wilczek:2000ih}. The latter is a running quantity which sets the 
strength of the strong interaction for all momenta. This running can be, a priori, inferred from the theory and 
encoded in the Renormalization Group equation (RGE) of $\alpha_s$, the value of which can be thus propagated from 
one given momentum to any other. The strong coupling is expressed by either the boundary condition for its RGE, 
generally dubbed $\Lambda_{\rm QCD}$, or its value at a reference scale, typically the $Z^0$-pole mass. 
This value is considered one of the QCD fundamental parameters, to be fitted from experiments, and amounts 
to $\alpha_s(m_Z^2) = 0.1181(11)$~\cite{Tanabashi:2018oca}, in the $\overline{\rm MS}$ renormalization scheme. Its current uncertainty 
of about 1 \% renders it the least precisely known of all fundamental coupling constants in nature. But at the 
same time, it is interesting to mention that a plethora of computations of LHC processes 
depend on an improved knowledge of $\alpha_s$ to reduce their theoretical uncertainties~\cite{Anastasiou:2015ema}. 
Especially in the Higgs sector, the uncertainty of $\alpha_s$ dominates that for the $H \to c\bar c,\,gg$ 
branching fractions and, after the error in the bottom mass, the one for the dominant $H \to b\bar b$ partial 
decay. And contrarily to other sources of uncertainty, as parton distribution functions, which reduced substantially~\cite{Ball:2018iqk}, 
that for $\alpha_s$ has not significantly changed in the last decade. Moreover, the $\alpha_s$ running and its 
uncertainty also has a non-negligible impact in the study of the stability 
of the electroweak vacuum, in the determination of the unification scale for the interaction couplings and, 
generally, in discriminating different New Physics scenarios.

There are many methods to determine the QCD coupling constant based on precision measurements of different processes and at different energy scales. A description of which can be found in the last QCD review of Particle Data Group (PDG)~\cite{Tanabashi:2018oca} or in specific reviews as, for instance, Ref.~\cite{Bethke:2011tr}. Alternatively, lattice QCD can be applied as a tool to convert a very precise physical observation, used for the lattice spacing setting, into $\Lambda_{\rm QCD}$. Thus, lattice QCD calculations can potentially be of a great help to increase the accuracy of our knowledge of $\alpha_s$. A review of most of the procedures recently implemented to determine the strong coupling from the lattice can be found in Ref.~\cite{Aoki:2016frl}. Among these procedures, there are those based on the computation of QCD Green's functions (see for instance~\cite{Alles:1996ka,Boucaud:2000ey,Boucaud:2001qz}), the most advantageous of which exploits the ghost-gluon vertex renormalized in the so called Taylor scheme~\cite{Sternbeck:2007br,Sternbeck:2010xu,Sternbeck:2012qs,Boucaud:2008gn,Blossier:2010ky,Blossier:2011tf,Blossier:2012ef,Blossier:2013ioa} such that the involved coupling can be computed from two-point Green's functions. As a bonus, this coupling has many phenomenological implications~\cite{Aguilar:2009nf,Aguilar:2010gm,Aguilar:2013xqa,Aguilar:2018csq,Aguilar:2018vfb} and is connected to the quark-gluon interaction kernel in continuum approaches to the QCD bound-state problem~\cite{Binosi:2014aea,Binosi:2016xxu,Binosi:2016nme,Rodriguez-Quintero:2018wma,Binosi:2018rht}. In this letter, we shall focus on this method and evaluate the Taylor coupling from lattice simulations with three Domain Wall fermions (DWF) at the physical point.  DWF (cf.~\cite{Vranas:2000tz, Kaplan:2009yg} for two interesting reviews), owing to their very good chiral properties, are expected to suffer less the impact of discretization artifacts.

%---------------

%\section{The running coupling from QCD two-point Green's functions}
%\label{sec:proc}
\smallskip
\noindent\emph{The running coupling from QCD two-point Green's functions}.\,---\,
First, we will sketch how the strong running coupling can be obtained from the gauge sector of QCD, invoking only 2-point Green's functions. Let $F$ and $D$ be the form factors (dressing functions) of the ghost and gluon propagators in the Landau gauge, the coupling will then read~\cite{Sternbeck:2007br,Boucaud:2008gn} 
%---------------------------------------
\beq\label{alpha} 
\alpha_T(k^2) \equiv \frac{g^2_T(k^2)}{4 \pi} = \lim_{a \to 0} 
\frac{g_0^2(a)}{4 \pi} F^{2}(k^2,a) D(k^2,a) \ ,
\eeq
%---------------------------------------
renormalized in the Taylor scheme, where $a$ stands for a regularization cut-off (the lattice spacing that is taken to vanish in the end of the calculation). 
% mentioning here where the connection of 2-point and 3-points coupling roots
The gauge-field 2-point Green's functions can be obtained from lattice QCD simulations with an extremely high level of accuracy, and combined next as \refeq{alpha} indicates to produce a precise estimate for the running of the coupling over a large window of momenta. Roughly above $3.5-4$ GeV, this running can be very well described by~\cite{Blossier:2010ky}:  
%--------------------------------------------------
\beq\label{alphahNP}
\alpha_T(k^2)
&=& 
\alpha^{\rm pert}_T(k^2)
\ 
\left( \rule[0cm]{0cm}{0.85cm}
 1 + \frac{9}{k^2} \
R\left(\alpha^{\rm pert}_T(k^2),\alpha^{\rm pert}_T(q_0^2) \right) \right.
\nonumber \\
&\times& \left. \left( \frac{\alpha^{\rm pert}_T(k^2)}{\alpha^{\rm pert}_T(q_0^2)}
\right)^{1-\gamma_0^{A^2}/\beta_0} 
\frac{g^2_T(q_0^2) \langle A^2 \rangle_{R,q_0^2}} {4 (N_C^2-1)}
\right) , %\nonumber \\
\eeq
%-----------------------------------------------------
where the perturbative result is supplemented by a leading Operator Product Expansion (OPE) non-perturbative correction, driven by the dimension-two gluon condensate $g^2_T(q_0^2) \langle A^2 \rangle_{R,q_0^2}$, including its anomalous dimension: $1-\gamma_0^{A^2}/\beta_0$ = 1/4 for $N_f$=3~\cite{Gracey:2002yt,Chetyrkin:2009kh} and 
%-------------------------------------
\beq
&&R\left(\alpha,\alpha_0\right) =
\left( 1 + 1.05882 \alpha + 1.16814 \alpha^2 + 1.95534 \alpha^3 \right) \nonumber \\ 
&&\times \left( 1 - 0.62446 \alpha_0 - 0.26140 \alpha_0^2 - 0.04275\alpha_0^3 \right) \ ,
%\nonumber \\ 
\eeq
%-------------------------------------
obtained here for $N_f$=3 as described in the Appendix of Ref.~\cite{Blossier:2010ky}. The momentum $q_0$=10 GeV is chosen as a subtraction point for the local operator. The perturbative $\alpha_T^{\rm pert}(k)$ can be approximated at the four-loop level by the integration of the $\beta$-function~\cite{Tanabashi:2018oca}, their coefficients being defined in the Taylor scheme~\cite{Chetyrkin:2000dq,Boucaud:2008gn}. Thus, the purely perturbative running reads as a function of $\ln{(k^2}/\Lambda_T^2)$, where $\Lambda_T$ stands for the $\Lambda_{\rm QCD}$-parameter in the Taylor scheme. The confrontation of lattice results, accurately obtained with \refeq{alpha}, to the running behavior predicted by \refeq{alphahNP} allows for a precise determination of the parameters $\Lambda_T$ and the gluon condensate $g^2_T(q_0^2) \langle A^2 \rangle_{R,q_0^2}$, both controlling the result displayed by \refeq{alphahNP}. Finally, as the running coupling in Taylor and $\overline{\rm MS}$ schemes relate as $\alpha_T = \overline{\alpha} ( 1 + c_1 \overline{\alpha} + {\cal O}(\overline{\alpha}^2))$, where $c_1$ is known~\cite{Chetyrkin:2000dq}, the $\Lambda_{\rm QCD}$-parameters can be in turn related, owing to their scale independence, by a subtraction of the couplings at asymptotically large momenta, thus obtaining~\cite{Blossier:2010ky}
%----------------------------------------
\beq\label{ratTMS}
\frac{\Lambda_{\overline{\rm MS}}}{\Lambda_T} \ = \ e^{\displaystyle -\frac{c_1}{2\beta_0}} \ = \ 
\exp{\left(\displaystyle - \frac{507-40 N_f}{792 - 48 N_f}\right)} 
\ .
\eeq
%----------------------------------------
All the procedure has been described in very detail in a series of articles, resulting from a long-term research program aimed at the determination of $\Lambda_{\overline{\rm MS}}$ from lattice QCD, where estimates for $N_f$=0~\cite{Boucaud:2008gn}, $N_f$=2~\cite{Blossier:2010ky} and $N_f$=2+1+1 (two degenerate light quarks and two non-degenerate ones with strange and charm flavors)~\cite{Blossier:2011tf,Blossier:2012ef,Blossier:2013ioa} have been delivered. 

The knowledge of $\Lms$ at a given $N_f$ defines the perturbative running of $\ams$, known to give a reliable effective description of the physical world between the energy thresholds of the $N_f$-th and ($N_f$+1)-th quark flavors for $N_f \ge$ 3~\cite{Tanabashi:2018oca}. 
Then, the matching formula 
%------------
\beq\label{eq:matching}
\ams^{N_f+1}(m_q) = \ams^{N_f}(m_q) 
\left( 1 + \sum_n c_{n0} \left(\ \ams^{N_f}(m_q)\right)^n \right) 
\eeq
%------------
can be applied to extend the running up to the threshold of the ($N_f$+2)-th quark flavor, where $m_q$ is the ${\overline{\rm MS}}$ running mass of the ($N_f$+1)-th quark and the coefficients $c_{n0}$ can be found in Ref.~\cite{Chetyrkin:2005ia} for $n\le$4. One can proceed this way up to the $Z^0$ mass scale. Thus,  the scale $\Lambda_T$, obtained for the running coupling in Taylor scheme at a given $N_f$, can be related to the benchmark value of $\ams(m_Z^2)$. 

%------------------------------------------

%\section{The running coupling at the physical point}
%\label{sec:lat}

\smallskip
\noindent\emph{The running coupling at the physical point}.\,---\,
Our previous determinations of $\Lms$ at $N_f<3$~\cite{Boucaud:2008gn,Blossier:2010ky} represented nothing but a heuristic effort, paving the way towards more realistic computations. This is so, first, because the lattice scale setting made by the confrontation with empiric observations is affected by the presence of the physical light flavors, up and down but also strange, thus inducing strong systematic effects. But moreover, even if one estimates and corrects the strange quark deviation in the $N_f$=2 case as prescribed in Ref.~\cite{Binosi:2016xxu}, the matching formula \eqref{eq:matching} can be hardly trusted at the strange-quark threshold. On the other hand, the one for $N_f$=2+1+1~\cite{Blossier:2011tf,Blossier:2012ef,Blossier:2013ioa} cannot be considered as a fully realistic estimate either, as far as it relies on lattice simulations where the lightest pseudoscalar mass ranges from 270 to 510 MeV and where chiral fits were required to take experimental $f_\pi$ and $m_\pi$ at the physical point~\cite{Baron:2010bv,Baron:2011sf}. %Additionally, all the previous studies were carried out with Twisted Mass fermions that despite being non-perturbatively $\mathcal{O}(a)$ improved should exhibit stronger discretization errors due to the explicit breaking of chiral symmetry. 

We repeat here the analysis with two ensembles of gauge-field configurations with 2+1 DWF simulated at the physical point and a third one with a pion mass of around 300 MeV (see Tab.~\ref{tab:setup}). Furthermore, we follow Ref.~\cite{Boucaud:2018xup} and  perform a very careful scrutiny of discretization lattice artifacts, which corresponds to taking $a \to 0$ in \refeq{alpha}, and approach thus the continuum limit. It can be outlined as follows: bare coupling and dressing functions are combined as \refeq{alpha} reads and $O(4)$-breaking artifacts cured by applying the $H(4)$-extrapolation~\cite{Becirevic:1999uc,Becirevic:1999hj,deSoto:2007ht}; residual $O(4)$-invariant artifacts are then removed by identifying ${\cal O}(a^2)$-corrections after a thorough comparison of results from the two different simulations at the physical point (as described in Sec.III.B of Ref.~\cite{Boucaud:2018xup}); and, finally, the outcome is checked by applying the same ${\cal O}(a^2)$-corrections to the third simulation's results. 

%---------------------------------------
%vskip -0.95 cm
\begin{figure}[t]
  \begin{center}
   \includegraphics[width=7.9cm]{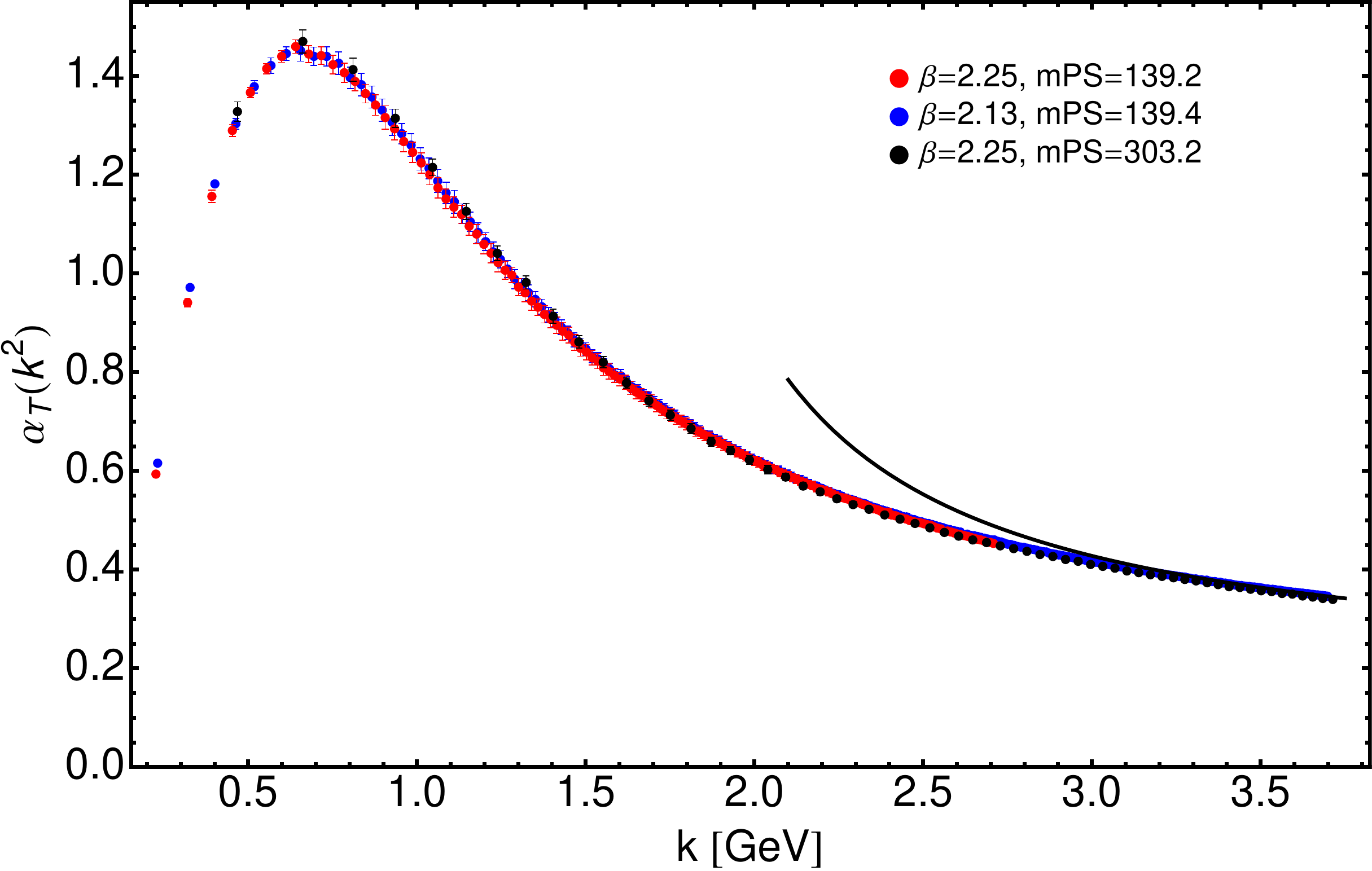}
  \end{center}
\vskip -0.5cm
\caption{\small 
running coupling data for the three lattice ensembles of Tab.~\ref{tab:setup}, after being cured for the discretization artifacts. The errors have been obtained by applying the jackknife method. The black solid line displays the result of \refeq{alphahNP} for $\Lambda_T$=581.5 MeV, corresponding to $\Lms$=320 MeV, and $g^2_T(q_0^2) \langle A^2 \rangle_{R,q_0^2}$=4.1 GeV$^2$.}
\label{fig:alphaT}
\end{figure}
%----------------------------------------

\begin{table}
  \caption{Set-up parameters for the $N_f$=2+1 ensembles exploited here~\cite{Allton:2007hx,Allton:2008pn,Arthur:2012yc,Blum:2014tka,Boyle:2017jwu}, which are generated with the Iwasaki gauge action~\cite{Iwasaki:1985we} and the DWF action~\cite{Kaplan:1992bt,Shamir:1993zy}. The two physical point ensembles use the M\"obius kernel~\cite{Brower:2004xi} while the heavier one uses the Shamir kernel~\cite{Kaplan:1992bt,Shamir:1993zy}.}
  \begin{center}
    \begin{tabular}{c|cccccc}
%      \hline
      $\beta$ & $L^3 \times T /a^4$ &  $a^{-1}[\mathrm{GeV}]$ & $m_\pi[\mathrm{MeV}$] & $m_\pi L $& V [fm$^4$]  & confs \\
      \hline
      2.13 &  $48^3 \times 96$    & 1.7295(38) &  139.4  & 3.9&5.47$^3\times$10.93 & 350\\
      2.25 &  $64^3 \times 128$   & 2.3586(70) &  139.2  & 3.8 &5.35$^3\times$10.70 &330\\
\hline
      2.25 & $32^3 \times 64$    & 2.3833(86) & 303.2  & 4.1& 2.65$^3\times$5.30 &330 \\
%      \hline
    \end{tabular}
  \end{center}
  \label{tab:setup}
\end{table}
%----

Figure~\ref{fig:alphaT} shows the coupling data for the three ensembles, exhibiting an excellent overlap and displaying a nice running behavior. Besides the good chiral properties of the DWF, a second ace of the exploited ensembles at the physical point is their large physical volume, which is made apparent by the absence of finite-volume effects when their results compare with those for the half-volume third ensemble. The upper bound of the running window, defined by $ka(2.25)=\pi/2$, corresponds to the largest lattice momenta which, being conservative, can be safely cured for discretization artifacts.  

\begin{figure}[t]
  \begin{center}
    \includegraphics[width=8.0cm]{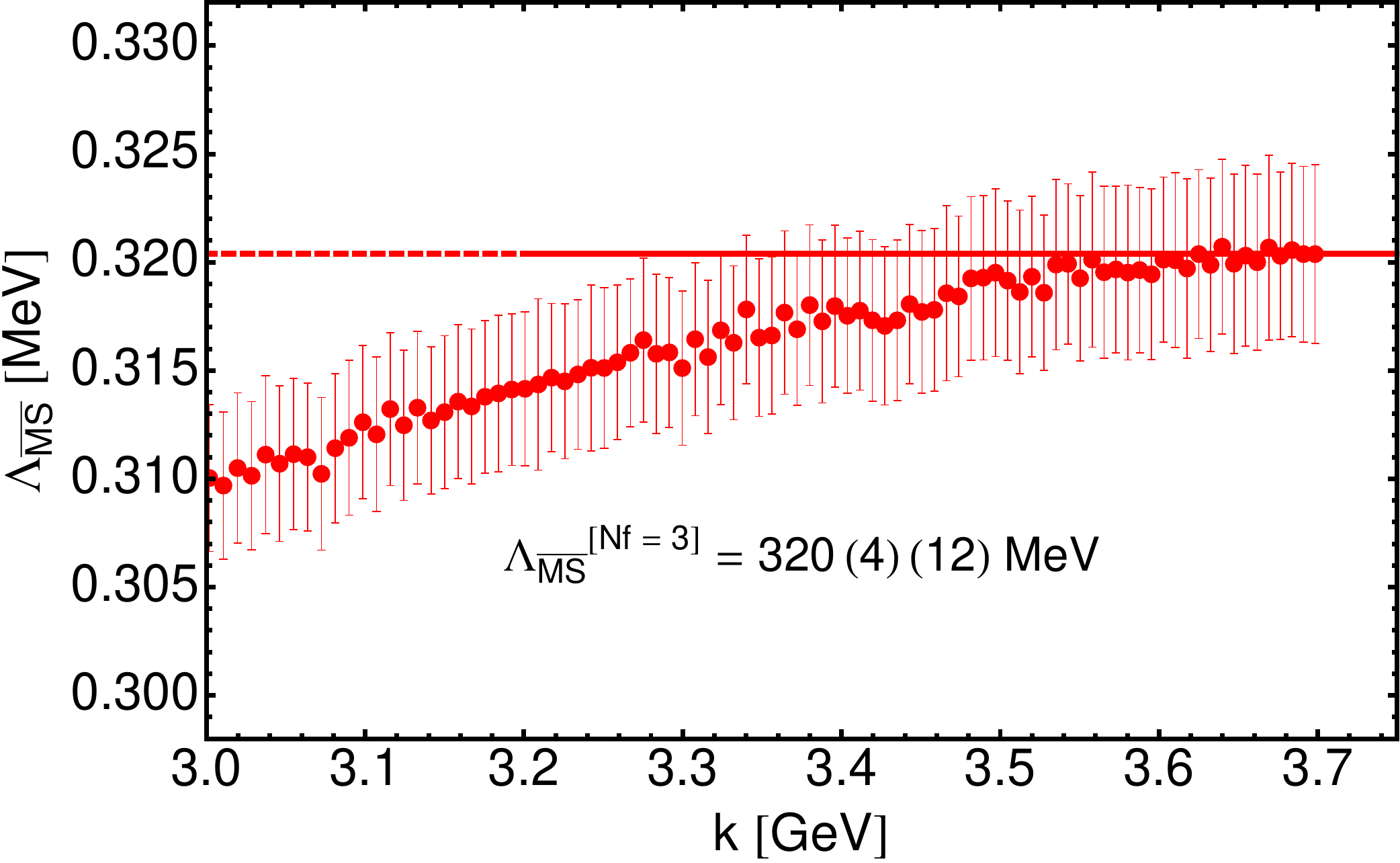}
  \end{center}
\vskip -0.5cm
\caption{\small Estimates of $\Lms$ (red solid circles) for all the lattice calculations of $\alpha_T(k^2)$ with $k > 3$ GeV, made through the numerical inversion of \refeq{alphahNP} with $g^2_T(q_0^2) \langle A^2 \rangle_{R,q_0^2}=4.1(1.1)$. The error bars displayed in the plot correspond  to the propagation of the uncertainty in the lattice determination of $\alpha_T$. 
}
\label{fig:lambda}
\end{figure}

The rightness of \refeq{alphahNP} and the need of the gluon condensate $g^2_T(q_0^2) \langle A^2 \rangle_{R,q_0^2}$ for the appropriate description of the momentum running of MOM-renormalized gauge-field Green's functions have been very well established~\cite{Boucaud:2000ey,Boucaud:2000nd,Boucaud:2001st,Boucaud:2001qz,Boucaud:2005xn,Boucaud:2008gn,Blossier:2010ky,Blossier:2011tf,Blossier:2012ef,Blossier:2013ioa,Boucaud:2013jwa}. Its nature and implications have been also thoroughly investigated in a exhaustive bunch of different analyses~\cite{Gubarev:2000nz,Boucaud:2002nc,RuizArriola:2004en,Dudal:2005na,Megias:2005ve,Boucaud:2005rm,RuizArriola:2006gq,Kondo:2006ih,Megias:2007pq,Megias:2009mp,Dudal:2010tf,Dudal:2011gd,Chang:2011mu,Boucaud:2011eh}, and its little dependence with the number of dynamical flavors found as well. Indeed, the effect of the heavier flavors can be thought to be negligible. Therefore, we have made the weighted-by-the-errors average of 2.7(1.0) and 4.5(5) GeV$^2$ for the gluon condensate at, respectively, $N_f$=2~\cite{Blossier:2010ky}\footnote{We have also considered the 6 \% of correction resulting from the strange quark in the lattice scale setting prescribed by \cite{Binosi:2016xxu}.} and $N_f$=4~\cite{Blossier:2012ef}; and thus finding for $N_f$=3: $g^2_T(q_0^2) \langle A^2 \rangle_{R,q_0^2}=4.1(1.1)$ GeV$^2$, where the uncertainty has been conservatively estimated by adding the errors in quadrature. We have then applied this value to \refeq{alphahNP} and inverted it numerically for 
all the lattice calculations at the physical point of $\alpha_T(k^2)$, with $k > 3$ GeV, and obtained thus the estimates of $\Lambda_T$, converted to $\Lms$ through \refeq{ratTMS} and displayed in Fig.~\ref{fig:lambda}. The plot shows a slow systematic decreasing below 3.62 GeV which, as proven in refs.~\cite{Blossier:2012ef,Blossier:2013ioa} for $N_f$=4, reflects that in this range higher-order nonperturbative corrections need to be included in \refeq{alphahNP}. Above 3.62 GeV, a small plateau appears: 11 points for which their central values differ as much as one per mil, their statistical errors being of the order of one per cent. However, the plateau is too small  to apply the same fitting strategy developed in refs.~\cite{Blossier:2012ef,Blossier:2013ioa}, with two free parameters and the lowest bound of the fitting window to be determined by the minimization of $\chi^2$. We make here no fit but take from literature the value of the condensate and evaluate $\Lms$ instead from the largest available momentum: $\Lms$=320(4)(13) MeV; the first quoted error results from propagating the one of $\alpha_T$ for this momentum into $\Lms$ determined by \refeq{alphahNP}, while the second one propagates the larger uncertainty from the condensate value. The central values for $\Lms$ and the condensate applied to \refeq{alphahNP} produce the black solid curve in Fig.~\ref{fig:alphaT}. 

%---------------------------------------
%vskip -0.95 cm
%----------------------------------------

Beyond this, one can also incorporate the same sort of higher-order nonperturbative correction effectively identified in Ref.~\cite{Blossier:2012ef} and try thus the same fit made therein. In so doing, one would obtain a nice plateau for momenta ranging from 2 to 3.7 GeV and 
a very consistent best-fit for $\Lms$=313 MeV, which will be used here only to estimate an uncertainty of 7 MeV resulting from the possible impact of higher-order nonperturbative corrections. Thus, 
following the matching procedure described in the previous section, we will be left with
%-----
\begin{equation}
\ams(m^2_Z) = 0.1172(3)(9)(5) \ , 
\end{equation}
%-----
where the first error propagates the uncertainty in the lattice determination of the Taylor coupling, the second does so for the value of the condensate and the last one stands for higher-order nonperturbative corrections.

%---------------------------------

%\section{$N_f$=2+1+1 versus $N_f$=2+1 results}

\noindent\emph{$N_f$=2+1+1 versus $N_f$=2+1 results}.\,---\,
Let us complete this analysis by relating the current results with our previous ones for $N_f$=2+1+1~\cite{Blossier:2012ef}. The lattice actions employed for the fermionic sector differ, twisted-mass for the latter and DWF for the former, although consistent results from both are expected in the continuum limit, if all discretization artifacts are indeed under control. The benchmark value of $\ams(m^2_Z)$ here is 
0.1172(11), all the errors combined in quadrature, and 0.1200(14) in Ref.~\cite{Blossier:2012ef}; both compatible with the current PDG world average~\cite{Tanabashi:2018oca}, 0.1181(11), but not with each other within their 1-$\sigma$ errors. This little difference might be due to a simple statistical deviation but can also reflect a small systematic effect in~\cite{Blossier:2012ef}, caused by the larger pion mass\footnote{The pion mass effect on the UV momentum running is seen to be very small in Fig.~\ref{fig:alphaT}, but effects on the physical scale setting and on the impact of the discretization artifacts cannot be excluded. Specially the latter would require a very accurate control of the continuum limit that might not have been achieved in Ref.~\cite{Blossier:2012ef}.}. This new updated result from the Taylor coupling, now at the physical point, lies closer to the FLAG lattice average\footnote{The FLAG average includes determinations from current two-point correlators~\cite{Bazavov:2014soa,Chakraborty:2014aca,McNeile:2010ji}, Schr\"odinger functional~\cite{Aoki:2009tf} and Wilson loops~\cite{Maltman:2008bx}}: 0.1182(12)~\cite{Aoki:2016frl}; but even closer to the non-lattice average of PDG: 0.1174(16)~\cite{Tanabashi:2018oca}. The PDG lattice unweighted average is in turn 0.1188(13), including the ghost-gluon determination~\cite{Blossier:2012ef,Blossier:2013ioa} among a few others~\cite{Bazavov:2014soa,Aoki:2009tf,Chakraborty:2014aca,McNeile:2010ji,Maltman:2008bx}. However, updating for the ghost-gluon with the current result, one is left with 0.1184(13), closer to the FLAG central value. 
%and, had we weighted with errors, both central values would have been the same: 0.1182. 
A very accurate $\ams(m^2_Z)=0.11852(84)$, obtained from the Sch\"odinger functional and renormalized couplings defined via the Gradient flow, has been also recently reported~\cite{Bruno:2017gxd}, with which our estimate agrees as well. It is noteworthy that this agreement demonstrates and strongly confirms, the approaches being radically different, that lattice systematics are well under control. In~\cite{Hudspith:2018bpz} DWF have also been employed for the extraction of $\alpha_s$ from the hadronic vacuum polarization function albeit not at the physical point. Their result is less precise, 0.1181(27), but anyhow in good agreement with ours.
%---------------------------------------
%vskip -0.95 cm
\begin{figure}[t]
  \begin{center}
    \includegraphics[width=8.0cm]{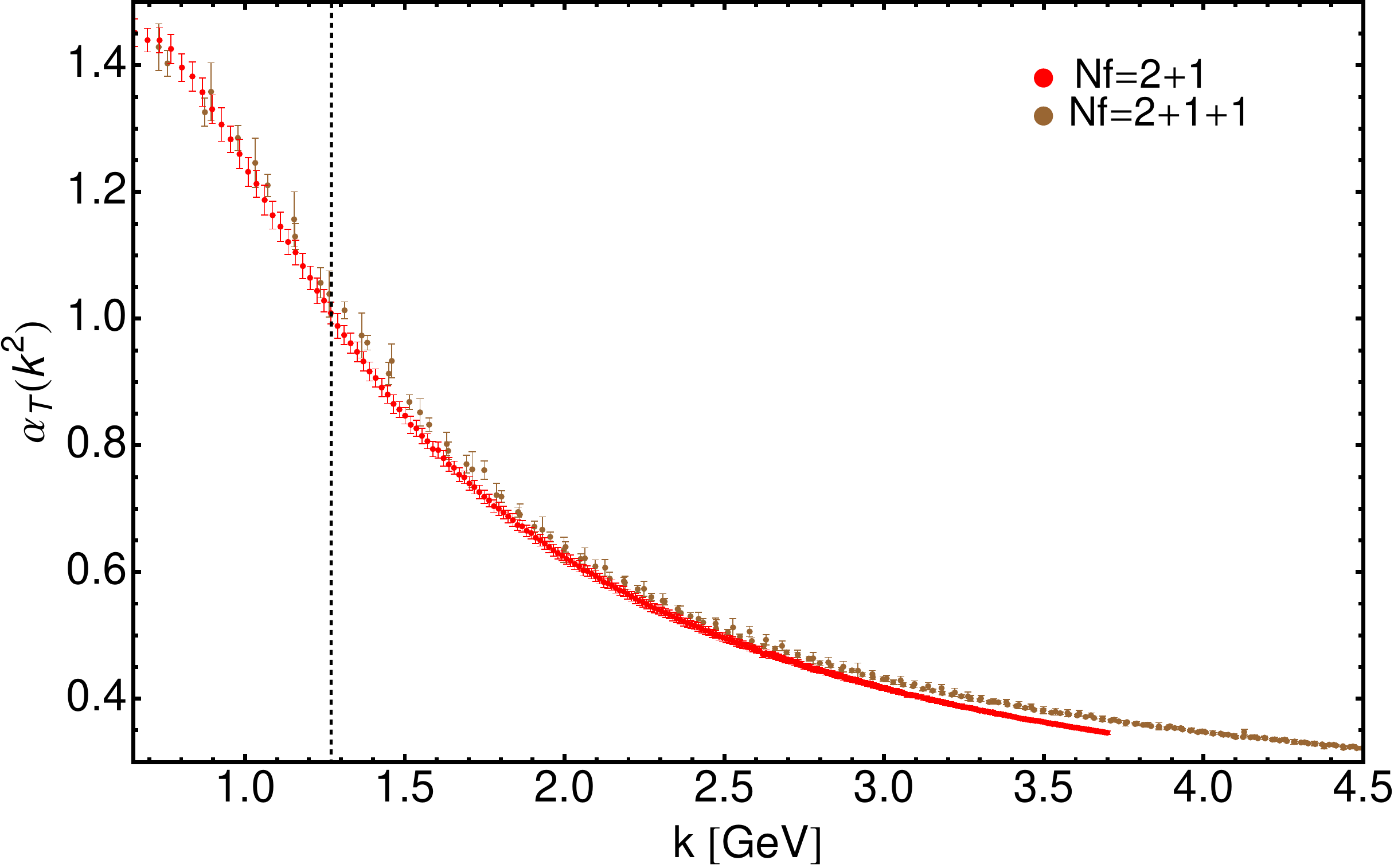}
  \end{center}
\vskip -0.5cm
\caption{\small Taylor coupling from the lattice, in the continuum limit, for $N_f$=2+1 DW dynamical flavors (red solid circles) and for $N_f$=2+1+1 twisted-mass flavors (brown solid circles) taken from Ref.~\cite{Blossier:2012ef}. The dotted line indicates the charm quark threshold at its $\overline{\rm MS}$ running mass. 
}
\label{fig:comp}
\end{figure}
%----------------------------------------

Beyond this, Fig.~\ref{fig:comp} displays a direct and striking comparison of the Taylor couplings for $N_f$=2+1 and 2+1+1. It is very apparent that, for momenta above the charm quark mass threshold, the 3-flavors coupling decreases faster than the 4-flavors one, extending down to nonperturbative momenta a well-known perturbative result: the beta function, the logarithmic derivative of the coupling with opposite sign, lessens when the number of flavors gets bigger. Around the charm threshold and below, within the deep IR domain, 4- and 3-flavors couplings appear to be the same, both reaching strikingly the same peak. Lightened by a few recent works~\cite{Binosi:2014aea,Binosi:2016xxu,Binosi:2016nme,Rodriguez-Quintero:2018wma,Binosi:2018rht}, bridging the gauge sector and phenomenological applications in QCD in connection with the bound-state problem, this feature can be well understood: the Taylor coupling can be related to the quark-gluon interaction kernel~\cite{Binosi:2016xxu}, both differing only by a small correction rooting in the ghost sector and not depending very much on the number of flavors. Fig.~\ref{fig:comp} thereby implies that the IR quark-gluon interaction strength does not depend on whether the charm quark becomes active or not.  Although expected, to our knowledge, this outcome has never been so remarkably exposed.

%---------------------------------

%\section{Conclusions} 

\noindent\emph{ Conclusions}.\,---\,
In this article we presented our results on the first computation of the strong running coupling at the physical point from the ghost-gluon vertex, computed from lattice simulations with $N_f$=2+1 DWF. We therewith update the last results~\cite{Blossier:2012ef,Blossier:2013ioa} obtained from applying the same procedure with four flavors but relatively large pion mass. The continuum limit treatment has been also herein improved. Thus, we have been left with an estimate for the benchmark value of $\ams(m^2_Z)$ more accurate, closer to the central value of the current lattice (FLAG) average~\cite{Aoki:2016frl} and in remarkably good agreement with the non-lattice average of PDG~\cite{Tanabashi:2018oca}. Moreover, lattice and non-lattice averages of PDG would come closer after updating the ghost-gluon determination. The convergence of lattice and non-lattice averages is very welcome, implying first that theory meets experiments but, not less important, that systematics effects from the discretization are under control and one can thereby take at face value the lattice errors, approaching thus the goal of getting the $\ams(m^2_Z)$ uncertainty below the one per cent level. 

On the other hand, the strong coupling in Taylor scheme, by itself, is an interesting quantity, as it can be directly related to the quark-gluon interaction kernel in continuum approaches to the QCD bound-state problem. It has been herein obtained for three dynamical quarks at the physical point, and has been shown to compare very well with previous results for four dynamical quarks but non-physical pion mass, qualitatively but also  quantitatively, beyond the small deviations due, presumably, to the larger pion mass and that impacts on the very delicate extraction of $\ams(m^2_Z)$. Such a comparison shows that the activation of the charm quark does not significantly affect the infrared quark-gluon interaction strength, and it only makes the running coupling decrease slower, above the charm threshold, as suggested by the $\beta$-function.

\noindent\emph{ Acknowledgements}.\,---\,
%\section*{Acknowledgements} 
We are indebted to the RBC/UKQCD collaboration, especially to Peter Boyle, Norman Christ, Zhihua Dong, Chulwoo Jung, Nicolas Garron, Bob Mawhinney and Oliver Witzel, for access to the lattices used in this work and to Rainer Sommer for fruitful discussions.
We thank the support of MINECO FPA2017-86380-P grant and SZ that of the DFG Collaborative Research Centre SFB 1225 (ISOQUANT). Numerical computations have used resources of CINES, GENCI IDRIS (project id 52271)  and of the IN2P3 computing facility in France.

%%%%%%%%%%%%%%%%%%%%%%%%%%%%%%%%%%%%%%%%%%%%%%%%%%%%%%%%%
%%                              BIBLIO
%%%%%%%%%%%%%%%%%%%%%%%%%%%%%%%%%%%%%%%%%%%%%%%%%%%%%%%%%

%\bibliographystyle{unsrt} % not to be used with revtex
\bibliography{total}

\end{document}